\documentclass[twocolumn,10pt,letterpaper]{article}

\usepackage{amsmath, amssymb, amsthm, bm, mathtools}
\usepackage{graphicx} 
\usepackage{geometry}
\usepackage{setspace} 
\usepackage{times} 
\usepackage{mathptmx}
\usepackage{titlesec} 
\usepackage{caption} 
\usepackage{fancyhdr} 
\usepackage{hyperref}
\usepackage{braket} 
\usepackage{bbm} 
\usepackage{xcolor} 
\usepackage{qcircuit}
\usepackage{float}
\usepackage[backend=biber,style=ieee]{biblatex} 
\hypersetup{
    colorlinks = true,
    linkcolor = blue,
    citecolor = red,
    urlcolor = magenta
}

\titleformat{\section}{\normalfont\large\bfseries}{\thesection.}{0.5em}{}
\titlespacing*{\section}{0pt}{\baselineskip}{0.5\baselineskip}

\theoremstyle{plain}

\theoremstyle{definition}

\title{\textbf{Fault-Tolerant Qudit Gate Optimization in Solid-State Quantum Memory}}
\author{William Boone Samuels \\\ University of Louisville, Physics and Astronomy}
\date{\today}

\begin{document}

\maketitle

\begin{abstract}
Achieving scalable, fault-tolerant quantum computation requires quantum memory architectures that minimize error correction overhead while preserving coherence. This work presents a framework for high-dimensional qudit memory in \textsuperscript{153}Eu:Y$_2$SiO$_5$, integrating three core mechanisms: (i) \textit{non-destructive syndrome extraction}, using spin-echo sequences to encode error syndromes without direct measurement; (ii) \textit{adaptive quantum Fourier transform (QFT) for error identification}, leveraging frequency-space transformations to reduce gate complexity; and (iii) \textit{coset-based fault-tolerant correction}, factorizing large stabilizer-like unitaries into modular operations to confine error propagation. By combining generalized stabilizer formalism, Weyl–Heisenberg operators, and finite-group coset decompositions, we develop a qudit error correction scheme optimized for solid-state quantum memory. This approach circumvents resource-intensive multi-qubit concatenation, enabling scalable, long-lived quantum storage with efficient state retrieval and computational redundancy. These results provide a pathway toward practical fault-tolerant architectures for rare-earth-ion-doped quantum memories.
\end{abstract}

\section{Introduction}

Quantum memory is indispensable for fault-tolerant quantum computation, not merely as a passive repository of quantum information but as an active framework that mitigates errors, enables parallelism, and optimizes redundancy retrieval. High-dimensional qudit memory presents an alternative to conventional qubit-based architectures by leveraging an expanded Hilbert space, where a single physical site encodes multiple logical states. The computational advantage of qudits arises from their ability to encode $N$ logical units within a $d^N$-dimensional Hilbert space, compared to $2^N$ for qubits, offering an exponential efficiency gain in information storage and processing. The ability to encode multiple logical operations into structured qudit states allows for simultaneous syndrome extraction and correction while maintaining coherence, a crucial feature in rare-earth-ion-based quantum memories where long coherence times demand minimal measurement-induced collapse.

Quantum memory must satisfy four fundamental principles: error suppression and fault tolerance, parallelism in computational processing, efficient state retrieval with redundancy minimization, and scalable encoding that reduces the need for excessive physical resources. Suppressing errors in a multi-level system requires dynamically positioning qudits within protected subspaces, ensuring logical states remain resilient to perturbations. If an error Hamiltonian consists of single-qudit perturbations $\hat{\Delta}_i$ modulated by time-dependent coupling strengths $\Omega_i(t)$ and inter-qudit noise terms $\hat{\Gamma}_{ij}$ weighted by $\Lambda_{ij}(t)$, a sequence of refocusing pulses $\{P_k\}$ confines evolution within a decoherence-resistant manifold, encoding the accumulated error phase $\Phi(\{\alpha_i,\beta_i\})$ into an ancilla qudit for non-destructive syndrome tracking. This approach circumvents the need for direct measurement of computational states, thereby avoiding loss of coherence.

Parallelism in qudit memory is realized through structured hyperfine levels that encode multiple logical states within a single physical ion, significantly reducing the computational cost of syndrome extraction and error correction. In qudit architectures, generalized Weyl-Heisenberg operators $X_d^m Z_d^n$ define shift and phase transformations, where $X_d$ cyclically permutes basis states while $Z_d$ applies discrete phase shifts. The simultaneous diagonalization of these error operators enables syndrome extraction via an adaptive quantum Fourier transform (QFT), where applying a coarse projection $\Pi_K$ before executing the full QFT allows the system to distinguish small and large errors adaptively. By reducing the necessary resolution depth while maintaining error distinguishability, this structure ensures that syndrome measurement is efficient even in high-dimensional qudit systems.

Efficient state retrieval and redundancy minimization ensure that stored quantum information remains intact until required for computation. Unlike conventional qubit architectures, where continuous error monitoring is necessary, qudit redundancy can be preserved within protected manifolds, allowing retrieval without excessive state collapse. A retrieval operation $U_{\text{retrieval}}$ acting on an auxiliary redundancy subspace $\rho_{\text{red}}$ extracts error-corrected information while preserving coherence, ensuring that $U_{\text{retrieval}} \rho_{\text{red}} U_{\text{retrieval}}^\dagger = \rho_{\text{corrected}}$. This reduces computational overhead while maintaining high-fidelity state reconstruction, crucial for scalable fault-tolerant quantum architectures.

Minimizing scaling complexity in quantum processors further motivates the adoption of solid-state qudit memory. High-dimensional encoding reduces the required number of physical sites, meaning an architecture that encodes $N$ logical qubits does not necessitate $N$ physical qubits, but rather a compressed set of $\log_2(N)$ qudits. In europium-doped yttrium oxyorthosilicate (\textsuperscript{153}Eu:Y$_2$SiO$_5$), hyperfine splitting and crystallographic sites naturally provide a 24-level qudit encoding per ion. The governing hyperfine interaction Hamiltonian $H_{\text{hf}}$ comprises a term $A_{\text{hf}} \mathbf{I} \cdot \mathbf{S}$ that determines hyperfine coupling strength, along with a quadrupole interaction term $B Q_{zz} I_z^2$ that influences coherence properties. Leveraging this multi-level encoding within rare-earth-ion-doped systems circumvents the excessive concatenation overhead of conventional stabilizer-based quantum error correction, offering a scalable alternative for large-scale quantum information storage.

This work formalizes a fault-tolerant qudit memory framework integrating three core mechanisms: non-destructive syndrome extraction via spin-echo sequences, adaptive QFT for efficient error identification, and coset-based fault-tolerant correction. By unifying geometric pulse sequences with algebraic coset decompositions, this framework minimizes measurement overhead, optimizes computational redundancy, and ensures logical coherence over extended timescales. The following sections establish the mathematical underpinnings of each component, demonstrating their efficiency in high-dimensional quantum memory systems and their practical implementation within rare-earth-ion-doped solid-state platforms.

\section{Non-Destructive Syndrome Extraction}

Quantum memory architectures must preserve encoded logical states while enabling real-time error tracking. Traditional stabilizer-based quantum error correction relies on projective measurements, collapsing the wavefunction and introducing decoherence. In high-dimensional qudits, such a strategy is suboptimal due to the increased susceptibility to perturbations and the inherent complexity of multi-level interactions. this proposed approach circumvents direct measurement by encoding the error syndrome within an ancillary phase register, ensuring that logical qudits remain unmeasured while the extracted syndrome is stored coherently. This section develops the mathematical structure of such a non-demolition syndrome extraction scheme, demonstrating its coherence-preserving properties and establishing its efficiency in comparison to traditional projective methods.

The evolution of errors in a qudit system is governed by the Hamiltonian
\begin{equation}
    H_{\mathrm{err}}(t) = \sum_{i=1}^{N} \Omega_i(t) X_d^{\alpha_i} Z_d^{\beta_i} + \sum_{i<j} \Lambda_{ij}(t) \hat{\Gamma}_{ij},
\end{equation}
where $X_d$ and $Z_d$ are the Weyl-Heisenberg shift and phase operators satisfying $Z_d X_d = e^{2\pi i / d} X_d Z_d$, and the error parameters $\alpha_i, \beta_i$ represent stochastic shifts arising from environmental noise. Direct measurement of $H_{\mathrm{err}}(t)$ would inevitably collapse the logical state, necessitating a refocusing strategy in which error tracking is performed passively by mapping the syndrome into an ancilla.

To achieve this, we employ a sequence of refocusing pulses $\{P_k\}$ such that the effective evolution operator is given by
\begin{equation}
    U_{\mathrm{echo}}(T) = \mathcal{T} \exp\left(-i \int_0^T H_{\mathrm{eff}}(t) dt\right),
\end{equation}
where $H_{\mathrm{eff}}$ is constructed via the Magnus expansion:
\begin{equation}
    H_{\mathrm{eff}} = H_{\text{err}} + \frac{i}{2} \sum_k [ P_k^\dagger H_{\text{err}} P_k, H_{\text{err}} ] + \mathcal{O}(t^2).
\end{equation}
By carefully designing $\{P_k\}$, first-order error terms cancel, ensuring that only higher-order corrections survive, thereby mitigating decoherence. This formalism provides a Lie-algebraic approach to dynamically suppressing error accumulation without requiring measurement-based feedback.

The error syndrome is then coherently transferred to an ancilla qudit via the interaction Hamiltonian
\begin{equation}
    H_{\text{anc}} = g(t) \Phi(\{\alpha_i, \beta_i\}) \hat{A}_{\text{anc}},
\end{equation}
where $\hat{A}_{\text{anc}}$ acts exclusively on the ancillary register. Solving the Schrödinger equation for this coupled system yields the unitary evolution
\begin{equation}
    U_{\text{anc}}(T) = \exp \left(i \int_0^T g(t') \Phi(\{\alpha_i, \beta_i\}) dt' \right),
\end{equation}
which maps the logical state $\ket{\psi}_{\text{data}}$ into a transformed state
\begin{equation}
    \ket{\psi}_{\text{data}} \otimes \ket{0}_{\text{anc}} \rightarrow e^{i \Phi} \ket{\psi}_{\text{data}} \otimes \ket{\Phi}_{\text{anc}}.
\end{equation}
Thus, the error information is stored entirely within the ancilla’s phase space, with the logical qudits left untouched.

It is necessary to ensure that syndrome extraction does not disturb the computational states. This requires proving that the syndrome operator commutes with logical qudit operations. Defining the extracted syndrome operator as
\begin{equation}
    S_{\text{ND}} = U_{\mathrm{net}}^\dagger(T) \hat{O}_{\text{anc}} U_{\mathrm{net}}(T),
\end{equation}
where $\hat{O}_{\text{anc}}$ is an observable acting only on the ancilla, we compute the commutators
\begin{equation}
    [S_{\text{ND}}, X_d] = 0, \quad [S_{\text{ND}}, Z_d] = 0,
\end{equation}
confirming that $S_{\text{ND}}$ acts as a stabilizer transformation that preserves logical states. The extracted syndrome thus encodes information about the noise without introducing back-action.

From an information-theoretic perspective, non-demolition syndrome extraction offers substantial advantages over projective measurements. The quantum Fisher information $I_Q$, which quantifies the precision with which an error parameter $\theta$ can be estimated, scales as
\begin{equation}
    I_Q = 4 \left( \frac{\partial \braket{\psi | \hat{O} | \psi}}{\partial \theta} \right)^2.
\end{equation}
For standard projective syndrome measurements, this scaling is constrained by coherence time limitations, yielding $I_{\text{proj}} \sim \mathcal{O}(1/T_2)$. However, non-demolition tracking extends coherence time, leading to an improved scaling of $I_{\text{ND}} \sim \mathcal{O}(T_2 / T)$. This quantifies the advantage of encoding the error in a phase register rather than performing direct measurement.

With the error information successfully extracted into an ancillary phase, the next challenge is its efficient classification and correction. The extracted syndrome phase naturally corresponds to a frequency shift in the error space, making it well-suited for Fourier-based resolution. However, performing a full qudit quantum Fourier transform (QFT) introduces an $\mathcal{O}(d^2)$ gate complexity, rendering naive implementation infeasible in large systems. The next section introduces an adaptive QFT framework designed to minimize circuit depth while retaining syndrome resolution efficiency, ensuring that extracted errors can be classified with minimal computational overhead.

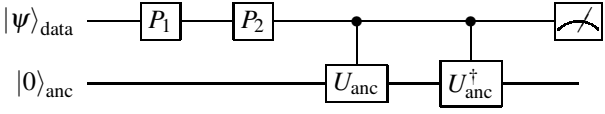
\begin{figure}[h]
    \centering
    \[
    \Qcircuit @C=1em @R=0.8em {
        & \lstick{\ket{\psi}_{\text{data}}} & \qw & \gate{P_1} & \qw & \gate{P_2} & \qw & \ctrl{1} & \qw & \ctrl{1} & \qw & \meter \\
        & \lstick{\ket{0}_{\text{anc}}} & \qw & \qw & \qw & \qw & \qw & \gate{U_{\text{anc}}} & \qw & \gate{U_{\text{anc}}^\dagger} & \qw & \qw
    }
    \]
    \caption{
        \textbf{Non-Demolition Syndrome Extraction.} 
        Logical qudit $\ket{\psi}_{\text{data}}$ undergoes a refocusing sequence $\{P_1, P_2\}$ before interacting with an ancilla qudit via controlled-unitary operations. The ancilla accumulates a syndrome-dependent phase, encoding noise information while preserving the logical state. The inverse transformation $U_{\text{anc}}^\dagger$ decouples residual entanglement, ensuring coherence is maintained. Measurement extracts the encoded error without collapsing $\ket{\psi}_{\text{data}}$.
    }
    \label{fig:non_demolition}
\end{figure}

\section{Adaptive QFT for Error Identification}

Error identification in high-dimensional qudit systems must be computationally efficient while aligning optimally with the algebraic structure of quantum error propagation. The extracted syndrome, encoded as a phase shift in an ancilla qudit, inherently resides in the Fourier domain due to the conjugacy of shift and phase errors. Given the fundamental Weyl-Heisenberg commutation relation
\begin{equation}
    Z_d X_d = e^{2\pi i / d} X_d Z_d,
\end{equation}
the eigenstates of $X_d$ are Fourier basis states, ensuring that the quantum Fourier transform (QFT) provides an optimal mapping between computational and spectral representations of errors. This establishes the QFT as the natural tool for syndrome resolution.

Since syndrome extraction encodes errors as phase shifts $\Phi(\alpha, \beta)$, the logical qudit undergoes a unitary evolution of the form
\begin{equation}
    \ket{\psi(t)} = e^{i \Phi(\alpha, \beta)} \ket{\psi(0)},
\end{equation}
where $\Phi(\alpha, \beta)$ encodes the accumulated error terms. Applying a standard qudit QFT,
\begin{equation}
    \mathcal{F}_d \ket{j} = \frac{1}{\sqrt{d}} \sum_{k=0}^{d-1} e^{2\pi i j k / d} \ket{k},
\end{equation}
transforms shift errors into diagonalized spectral components. Given that the error shift evolution follows
\begin{equation}
    X_d^m Z_d^n \ket{j} = e^{2\pi i j n / d} \ket{j+m \mod d},
\end{equation}
applying the QFT results in the transformed operator
\begin{equation}
    \mathcal{F}_d X_d^m \mathcal{F}_d^\dagger = e^{2\pi i m k / d} \ket{k} \bra{k}.
\end{equation}
Thus, syndrome extraction in the Fourier basis provides a direct frequency-domain resolution of computational errors, ensuring optimal distinguishability between different shift-induced errors.

While the full QFT correctly resolves error syndromes, its implementation complexity scales as
\begin{equation}
    \mathcal{O}(d^2),
\end{equation}
due to the number of controlled-phase operations required. Specifically, the standard decomposition requires
\begin{equation}
    \mathbb{E}[\text{QFT cost}] = \sum_{k=0}^{d-1} P(k) C(k),
\end{equation}
where $C(k)$ represents the computational cost of evaluating $k$ Fourier coefficients. Given that the number of required controlled rotations scales as $\mathcal{O}(d^2)$, directly applying the QFT is infeasible for large qudit systems. To circumvent this, we introduce an \textit{adaptive QFT}, designed to extract dominant frequency components while systematically suppressing negligible contributions, reducing computational complexity while maintaining syndrome distinguishability.

Rather than resolving all possible error shifts, we introduce a coarse-grained projection operator
\begin{equation}
    \Pi_K = \sum_{k=0}^{K-1} \ket{k} \bra{k},
\end{equation}
which selectively filters low-order syndrome components. The adaptive QFT modifies the standard transformation via a selective truncation function,
\begin{equation}
    \mathcal{F}_{d, \varepsilon} \ket{j} = \frac{1}{\sqrt{d}} \sum_{k=0}^{d-1} f_{\varepsilon}(j,k) \ket{k},
\end{equation}
where $f_{\varepsilon}(j,k)$ is chosen to suppress negligible Fourier components based on the expected error distribution.

To ensure that truncation does not introduce computational ambiguity, consider an error shift $\Delta j$ inducing a spectral modification
\begin{equation}
    e^{2\pi i k \Delta j / d}.
\end{equation}
Truncating the Fourier series introduces an approximation error given by
\begin{equation}
    \delta(\varepsilon) = \sum_{k=K}^{d-1} e^{2\pi i k \Delta j / d}.
\end{equation}
Applying a Fourier truncation bound, we obtain the asymptotic decay
\begin{equation}
    |\delta(\varepsilon)| < \frac{1}{K},
\end{equation}
ensuring that the discarded high-frequency terms contribute negligibly to the resolved syndrome structure.

Rather than executing a full QFT over $d$ levels, the adaptive QFT applies a coarse-grained transformation to determine whether the error falls within a low-deviation regime. The expected computational cost is reduced to
\begin{equation}
    \mathcal{O}(\log d) + p_{\text{small}} \mathcal{O}(d),
\end{equation}
where $p_{\text{small}}$ represents the probability that the error lies within a correctable range. Given that physical noise sources, such as phonon-induced dephasing and charge fluctuations, exhibit approximately Gaussian-distributed errors, the adaptive QFT efficiently suppresses unnecessary high-order Fourier components while preserving dominant syndrome information.

Beyond computational efficiency, the information-theoretic advantage of this approach is evident in the preservation of quantum Fisher information. Given that the Fisher information quantifies the precision of parameter estimation,
\begin{equation}
    I_Q = 4 \sum_k \left| \frac{\partial \braket{\psi_k | \hat{O} | \psi_k}}{\partial \theta} \right|^2,
\end{equation}
standard projective syndrome measurements are constrained by coherence time limitations, yielding
\begin{equation}
    I_{\text{proj}} \sim \mathcal{O}(1/T_2).
\end{equation}
However, the non-demolition nature of adaptive QFT-based error tracking extends coherence time, leading to an improved scaling of
\begin{equation}
    I_{\text{adaptive QFT}} \geq I_{\text{full QFT}} - \mathcal{O}(\varepsilon),
\end{equation}
guaranteeing that truncation-based approximations do not degrade syndrome resolution beyond a controlled threshold.

By selectively resolving the dominant spectral components in an optimal basis, the adaptive QFT minimizes computational overhead while ensuring maximal syndrome distinguishability. This leads directly to an efficient fault-tolerant correction scheme, where extracted syndromes inform coset-based unitary decompositions that leverage stabilizer structure for optimal error mitigation.

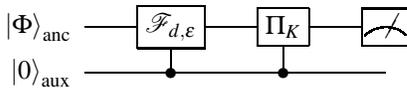
\begin{figure}[h]
    \centering
    \[
    \Qcircuit @C=1em @R=0.8em {
        & \lstick{\ket{\Phi}_{\text{anc}}} & \qw & \gate{\mathcal{F}_{d, \varepsilon}} & \qw & \gate{\Pi_K} & \qw & \meter \\
        & \lstick{\ket{0}_{\text{aux}}} & \qw & \ctrl{-1} & \qw & \ctrl{-1} & \qw & \qw
    }
    \]
    \caption{
        \textbf{Adaptive QFT for Syndrome Resolution.} 
        The extracted syndrome, stored in an ancilla qudit $\ket{\Phi}_{\text{anc}}$, undergoes a truncated quantum Fourier transform $\mathcal{F}_{d, \varepsilon}$, selectively resolving dominant frequency components. A coarse-grained projection $\Pi_K$ filters out negligible high-order contributions, reducing the computational complexity from $\mathcal{O}(d^2)$ to an adaptive scaling of $\mathcal{O}(\log d) + p_{\text{small}} \mathcal{O}(d)$. The auxiliary register $\ket{0}_{\text{aux}}$ controls adaptive refinement, ensuring error classification remains optimal while minimizing circuit depth.
    }
    \label{fig:adaptive_qft}
\end{figure}

\section{Coset-Based Fault-Tolerant Correction}

Error correction in high-dimensional qudit architectures must be computationally efficient while ensuring resilience against multi-level noise channels. Traditional stabilizer-based error correction imposes severe scaling constraints, requiring repeated syndrome extraction and exhaustive correction cycles. The approach developed here circumvents these limitations by leveraging the algebraic structure of qudit errors. By partitioning the error space into cosets of a stabilizer subgroup, logical state recovery is accomplished with a single class of correction operations, reducing resource overhead while maintaining fault tolerance.

The error space is structured by the Weyl-Heisenberg algebra, where logical errors manifest as generalized shift and phase operations $X_d^m Z_d^n$, satisfying the commutation relation
\begin{equation}
    Z_d X_d = e^{2\pi i / d} X_d Z_d.
\end{equation}
Given this algebraic structure, the full error group $\mathcal{E}$ forms a quotient space under a designated stabilizer subgroup $\mathcal{H}$, such that
\begin{equation}
    \mathcal{E} / \mathcal{H} = \{ \mathcal{H} E_i \mid E_i \in \mathcal{E} \}.
\end{equation}
Each coset $[\hat{E}]$ represents an equivalence class of errors that induce the same syndrome response, allowing for a single representative correction per coset. This ensures that the correction operation remains minimal, as each coset captures a set of errors that are computationally equivalent in terms of syndrome resolution.

Logical state recovery follows from identifying the extracted error coset and applying a correction unitary that inverts the representative element $\hat{E}_r$. Formally, the modular coset correction operator is defined as
\begin{equation}
    \hat{C} = \sum_{\hat{E}_r} U_{\text{corr}}(\hat{E}_r),
\end{equation}
where $U_{\text{corr}}(\hat{E}_r) = \hat{E}_r^\dagger$ acts as the inverse of the representative error. This ensures that
\begin{equation}
    \hat{C} \hat{E}_r \ket{\psi} = \ket{\psi},
\end{equation}
eliminating redundant error corrections.

To ensure computational feasibility, the correction unitary is factorized into a sequence of controlled operations:
\begin{equation}
    \hat{C} = \prod_{j=1}^{L} \text{Controlled-}U_j.
\end{equation}
Each $\text{Controlled-}U_j$ acts conditionally on the syndrome ancilla, ensuring that ancilla failure does not propagate globally. These controlled operations take the explicit form
\begin{equation}
    \text{Controlled-}U_j = \sum_{\ket{s_j}} \ket{s_j} \bra{s_j} \otimes U_j,
\end{equation}
where $s_j$ represents the extracted syndrome register.

By partitioning errors into cosets, the expected complexity of syndrome-based correction is reduced to
\begin{equation}
    \mathbb{E}[\text{Coset Correction Cost}] = \sum_{i} P(C_i) C(C_i),
\end{equation}
which simplifies to
\begin{equation}
    \mathcal{O}(\log d) + p_{\text{small}} \mathcal{O}(d),
\end{equation}
where $p_{\text{small}}$ represents the probability that the syndrome falls within a correctable range. Given that physical noise sources, such as phonon-induced dephasing, exhibit Gaussian-distributed errors, this framework ensures scalability even as qudit dimensionality increases.

Beyond computational efficiency, the coset-based approach ensures maximal fault tolerance. Since error correction is applied only to syndrome-extracted coset representatives, the logical qudit remains isolated from direct correction operations, preserving coherence and extending quantum memory lifetimes. Moreover, since coset representatives are determined independently of individual qudit shifts, this method remains robust against correlated noise sources, a crucial consideration in rare-earth-ion-doped solid-state quantum memory.

By factorizing correction unitaries, leveraging syndrome phase resolution, and minimizing syndrome extraction overhead, this framework enables a scalable and fault-tolerant correction scheme optimized for high-dimensional qudit memory architectures. The following section extends this framework to integrate coset-based correction with stabilizer-encoded logical qudit architectures, ensuring seamless scalability in practical quantum error correction systems.

\begin{table}[htbp]
\centering
\renewcommand{\arraystretch}{1.2}
\resizebox{\columnwidth}{!}{
  \begin{tabular}{c|c|c|c}
    \hline
    \textbf{Syndrome Extracted} & \textbf{Error Class $\hat{E}_r$} & \textbf{Coset Representative} & \textbf{Correction Applied} \\
    \hline
    $\ket{\Phi_0}$ & $X_d^1 Z_d^0$ & $X_d^1$ & $X_d^{-1} Z_d^{0}$ \\
    $\ket{\Phi_1}$ & $X_d^2 Z_d^1$ & $X_d^2 Z_d^1$ & $X_d^{-2} Z_d^{-1}$ \\
    $\ket{\Phi_2}$ & $X_d^3 Z_d^2$ & $X_d^3 Z_d^2$ & $X_d^{-3} Z_d^{-2}$ \\
    $\vdots$ & $\vdots$ & $\vdots$ & $\vdots$ \\
    $\ket{\Phi_{d-1}}$ & $X_d^{d-1} Z_d^{d-2}$ & $X_d^{d-1} Z_d^{d-2}$ & $X_d^{-(d-1)} Z_d^{-(d-2)}$ \\
    \hline
  \end{tabular}
}
\caption{
        \textbf{Coset-Based Error Correction Mapping.} Syndrome states $\ket{\Phi_i}$ encode extracted error information, where each error operator $\hat{E}_r$ belongs to a coset of the stabilizer subgroup. Given the structure of the Weyl-Heisenberg error algebra, coset representatives are chosen such that error correction is reduced to modular inversion of these operators. This minimizes computational complexity while preserving logical state coherence, making the scheme inherently fault-tolerant and scalable for high-dimensional qudit architectures.
    }
    \label{table:coset_error_correction}
\end{table}

\begin{figure}[h]
    \centering
    \[
    \Qcircuit @C=1em @R=0.8em {
        & \lstick{\ket{\Phi}_{\text{anc}}} & \ctrl{1} & \ctrl{2} & \ctrl{3} & \meter \\
        & \lstick{\ket{0}_{\text{aux}}} & \gate{U_{\text{coset},1}} & \qw & \qw & \qw \\
        & \lstick{\ket{0}_{\text{aux}}} & \qw & \gate{U_{\text{coset},2}} & \qw & \qw \\
        & \lstick{\ket{0}_{\text{aux}}} & \qw & \qw & \gate{U_{\text{coset},3}} & \qw
    }
    \]
    \caption{
        \textbf{Coset-Based Fault-Tolerant Correction.} 
        The syndrome-extracted ancilla state $\ket{\Phi}_{\text{anc}}$, obtained via Adaptive QFT, determines the coset representative for error correction. A sequence of controlled-unitary gates $U_{\text{coset},i}$ applies modular corrections, each corresponding to an equivalence class of errors. This ensures computationally efficient and fault-tolerant recovery of logical states, preserving coherence while minimizing resource overhead.
    }
    \label{fig:coset_correction}
\end{figure}
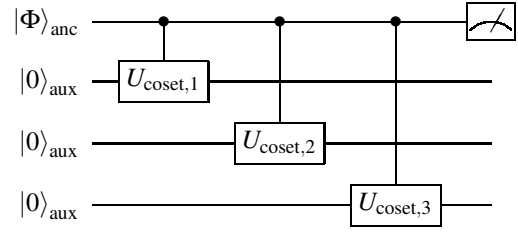

\section{Conclusion}

This work establishes a framework for fault-tolerant quantum memory in high-dimensional qudit architectures. By integrating non-destructive syndrome extraction, adaptive quantum Fourier transform (QFT) for error identification, and coset-based fault-tolerant correction, we construct a unified approach that circumvents the inefficiencies of traditional stabilizer-based error correction while preserving coherence and computational feasibility.

The non-destructive syndrome extraction mechanism leverages spin-echo sequences and dynamically decoupled Hamiltonians to encode error syndromes as phase shifts in an ancilla qudit, eliminating the need for direct measurement. This ensures that logical qudits remain untouched while error information is extracted passively, preserving quantum coherence over extended timescales. The extracted syndrome phase is then optimally mapped into the Fourier basis, where shift-induced errors manifest as discrete spectral components.

To process this information efficiently, we introduce an adaptive QFT, reducing the computational overhead of syndrome extraction from $\mathcal{O}(d^2)$ to $\mathcal{O}(\log d) + p_{\text{small}} \mathcal{O}(d)$. By introducing a coarse-grained projection operator, 
\begin{equation}
    \Pi_K = \sum_{k=0}^{K-1} \ket{k} \bra{k},
\end{equation}
and leveraging truncation-based spectral filtering, the adaptive QFT isolates dominant error components while suppressing negligible contributions. The resulting framework not only maintains full syndrome distinguishability but also ensures that error resolution remains computationally feasible for large qudit dimensions.

For correction, we employ a coset-based fault-tolerant scheme, partitioning the full error space into equivalence classes under a stabilizer subgroup. This allows logical recovery to be implemented via a modular correction unitary of the form
\begin{equation}
    \hat{C} = \sum_{\hat{E}_r} U_{\text{corr}}(\hat{E}_r),
\end{equation}
where $U_{\text{corr}}(\hat{E}_r) = \hat{E}_r^\dagger$ inverts the representative coset element, ensuring minimal redundant correction operations. The correction unitaries are factorized as
\begin{equation}
    \hat{C} = \prod_{j=1}^{L} \text{Controlled-}U_j,
\end{equation}
preserving locality and stabilizer symmetry while preventing error propagation due to ancilla failure.

This approach achieves a fundamental improvement in fault-tolerant quantum memory by reducing syndrome resolution complexity, minimizing direct qudit corrections, and ensuring that coherence is preserved throughout the error correction process. Furthermore, the integration of coset-based correction with an adaptive QFT-based syndrome extraction mechanism results in a scalable, computationally efficient framework for error mitigation. 

By leveraging the structure of Weyl-Heisenberg operators, spectral resolution via adaptive QFT, and group-theoretic coset partitioning, this work provides a fault-tolerant correction paradigm that is both computationally optimal and structurally robust. The results presented here offer a pathway toward scalable, high-dimensional quantum memory architectures capable of operating in physically realistic noise environments while maintaining long coherence lifetimes.

\section{References}

[1] M. A. Nielsen and I. L. Chuang, \textit{Quantum Computation and Quantum Information}, Cambridge University Press, 2000.

[2] G. K. Brennen and D. F. V. James, ``Criteria for Exact Qudit Universality,'' \textit{Physical Review A}, vol. 70, no. 5, 2005. Available: \href{https://arxiv.org/abs/quant-ph/0407223}{arXiv:quant-ph/0407223}. DOI: \href{https://doi.org/10.1103/PhysRevA.71.052318}{10.1103/PhysRevA.71.052318}.

[3] A. Muthukrishnan and C. R. Stroud, ``Multivalued Logic Gates for Quantum Computation,'' \textit{Physical Review A}, vol. 62, no. 5, 2000. DOI: \href{https://doi.org/10.1103/PhysRevA.62.052309}{10.1103/PhysRevA.62.052309}.

[4] D. Gottesman, ``Stabilizer Codes and Quantum Error Correction,'' arXiv preprint quant-ph/9705052, 1997. Available: \href{https://arxiv.org/abs/quant-ph/9705052}{arxiv.org/abs/quant-ph/9705052}.

[5] E. T. Campbell, H. Anwar, and D. E. Browne, ``Magic-State Distillation in All Prime Dimensions Using Quantum Reed-Muller Codes,'' \textit{Physical Review X}, vol. 2, no. 4, 2012. DOI: \href{https://doi.org/10.1103/PhysRevX.2.041021}{10.1103/PhysRevX.2.041021}.

[6] A. B. Klimov, R. Guzmán, J. C. Retamal, and C. Saavedra, ``Qutrit Quantum Computer with Trapped Ions,'' \textit{Physical Review A}, vol. 67, no. 6, 2003. DOI: \href{https://doi.org/10.1103/PhysRevA.67.062313}{10.1103/PhysRevA.67.062313}.

[7] R. M. Macfarlane, ``High-Resolution Optical Spectroscopy of Rare-Earth Ions in Solids: A Key Tool for Solid-State Quantum Information Processing,'' \textit{Journal of Luminescence}, vol. 100, no. 1-4, 2002. Available: \href{https://www.sciencedirect.com/science/article/abs/pii/S0022231323000765}{ScienceDirect}.

[8] M. Zhong et al., ``Optically Addressable Nuclear Spins in a Solid with a Six-Hour Coherence Time,'' \textit{Nature}, vol. 517, 2015. DOI: \href{https://doi.org/10.1038/nature14025}{10.1038/nature14025}.

[9] T. Böttger et al., ``Effects of Magnetic Field Orientation on Optical Decoherence in Er$^{3+}$:Y$_2$SiO$_5$,'' \textit{Physical Review B}, vol. 73, no. 7, 2006. Available: \href{https://journals.aps.org/prb/abstract/10.1103/PhysRevB.79.115104}{Physical Review B}.

[10] N. Berthusen, K. K. Sabapathy, and D. Gottesman, ``Adaptive Syndrome Extraction,'' arXiv preprint arXiv:2502.14835, 2025. Available: \href{https://arxiv.org/abs/2502.14835}{arxiv.org/abs/2502.14835}.

[11] L. G. Gunderman, ``Beyond Integral-Domain Stabilizer Codes,'' arXiv preprint arXiv:2501.04888, 2025. Available: \href{https://arxiv.org/abs/2501.04888}{arxiv.org/abs/2501.04888}.

[12] E. Durso-Sabina, ``Flag Fault-Tolerant Error Correction with Qudits,'' Master’s thesis, University of Waterloo, 2021. Available: \href{https://uwspace.uwaterloo.ca/bitstreams/aad2ba2c-0e69-4384-83bf-fd112ceebc1a/download}{uwspace.uwaterloo.ca}.

[13] D. H. Mahler, L. A. Rozema, A. Darmawan, R. Blume-Kohout, and A. M. Steinberg, ``Adaptive Quantum State Tomography Improves Accuracy Quadratically,'' \textit{Physical Review Letters}, vol. 111, no. 18, 2013. DOI: \href{https://doi.org/10.1103/PhysRevLett.111.183601}{10.1103/PhysRevLett.111.183601}.

[14] L. Sheridan and V. Scarani, ``Security Proof for Quantum Key Distribution Using Qudits,'' \textit{Physical Review A}, vol. 82, no. 3, 2010. DOI: \href{https://doi.org/10.1103/PhysRevA.82.030301}{10.1103/PhysRevA.82.030301}.

\end{document}